# Mimivirus and the emerging concept of « giant » virus

Jean-Michel Claverie[1,2+], Hiroyuki Ogata[1], Stéphane Audic[1,2], Chantal Abergel[1], Pierre-Edouard Fournier[1], Karsten Suhre[1]

[1] Information Génomique et Structurale, CNRS UPR 2589,
Institute of Microbiology and Structural Biology,
31 Chemin Joseph Aiguier, 13402 Marseille Cedex 20

[2] Faculté de Médecine, Université de la Méditerranée,
27 Blvd Jean Moulin, 13385 Marseille Cedex 5, France
Tel : (33) 491 16 45 48
Fax: (33) 491 16 45 49

[+] correspondance to:
E-mail : Jean-Michel.Claverie@igs.cnrs-mrs.fr

## Summary

The recently discovered *Acanthamoeba polyphaga* Mimivirus is the largest known DNA virus. Its particle size (>400 nm), genome length (1.2 million bp) and large gene repertoire (911 protein coding genes) blur the established boundaries between viruses and parasitic cellular organisms. In addition, the analysis of its genome sequence identified new types of genes not expected to be seen in a virus, such as aminoacyl-tRNA synthetases and other central components of the translation machinery. In this article, we examine how the finding of a giant virus for the first time overlapping with the world of cellular organisms in terms of size and genome complexity might durably influence the way we look at microbial biodiversity, and force us to fundamentally revise our classification of life forms. We propose to introduce the word "girus" to recognize the intermediate status of these giant DNA viruses, the genome complexity of which make them closer to small parasitic prokaryotes than to regular viruses.

## Introduction

The discovery of *Acanthamoeba polyphaga* Mimivirus (La Scola *et al.*, 2003) and the analysis of its complete genome sequence (Raoult *et al.* 2004) sent a shock wave through the community of virologists and evolutionists. The size, gene content, and phylogenetic characterization of the virus genome, challenged many accepted ideas about what virus should look like, and where they might come from. Several comments have already been published on Mimivirus (Ghedin and Fraser, 2005; Ghedin and Claverie, 2005; Desjardins et al., 2005; Koonin, 2005; Galperin 2005; Moreira and López-Garcia, 2005; Ogata et al., 2005a; Ogata et al., 2005b). However, the consequences of the qualitative and quantitative gaps separating it from previously known DNA viruses are yet to be analyzed in depth.

Very large DNA virus genomes have accumulated steadily in the databases, since the spectacular achievement of Barrell's team sequencing the 230 kb of human cytomegalovirus (Human herpesvirus 5) as early as 1990 (Chee et al., 1990). Curiously, however, these incremental progresses failed to generate much emotion or trigger significant changes in the perception/notion of virus that prevails in the general community of biologists. In our collective subconscious mind, viruses are still thought of as highly optimized minimal "bags of genes", packaging just enough information to deal with host infection and to highjack the host machinery for multiplying viral particles. Given the simplicity of a minimal particle (a capsid protein and a few more proteins for genome packaging), a viral genome is thus expected to carry less than a dozen of genes. In this context, a virus (or a phage) packing more than 300 genes already appears as an evolutionary absurdity, an "overkill".

Thus, if Mimivirus deserved some special attention, it is not primarily because it was larger than the previously largest virus before it; It was because it is the first virus the various dimensions of which (particle size and genome complexity) are significantly overlapping with those typical of parasitic cellular microorganisms (Table 1). With this unique feat, Mimivirus forces us to abandon our traditional size criteria, and prompt us to re-formulate a correct answer to the fundamental question: what is a virus?

This question is not only philosophical, or related to speculative thinking about the origin of life. It has very practical implications. Particle size, for instance, was always central to virus isolation protocols, and still directly pertains to the design of the modern "metagenomic" studies aiming at assessing microbial biodiversity. Simply acknowledging the fact that all virus might not be filterable through the typical "sterilizing" 0.2 – 0.3 µm-pore filters, already changes our interpretation of the currently available data, and call for significant changes to the protocol of future environmental sampling campaigns. The first part of this review article will focus on the size distribution of DNA viruses and propose that the largest of them might constitute a new type of microbial organisms, submitted to their own, yet unknown, peculiar evolutionary constraints.

The interpretation of metagenomic data is also seriously challenged by the genome complexity exhibited by these increasingly large DNA viruses, and the qualitative overlap of their gene contents with the ones of cellular organisms. Given the large number of their genes not obviously related to DNA replication and particle synthesis (many type of enzymes, components of signaling pathways, tRNAs, transcription and translation factors, …, etc), what criteria now remain at our disposal to reliably distinguish, using their sequences, viral genes from those belonging to the genome of a cellular organism? Those questions are addressed in the second part of this article.

**Table 1. Genome sizes of the largest viruses and of the smallest cellular organisms.**

Mimivirus genome is larger than the one of 25 cellular organisms. These organisms are obligate parasites or symbionts, with the exception of *T. whipplei* for which axenic growth conditions have recently been described (Renesto et al., 2003). Only publicly available genome sequences are taken into account. The size indicated for Bacillus page G refers to the unique part of the genome sequence. The total packaged DNA is about 650 kb long. The coding density is roughly same and about one protein coding gene/1 kb for all the above organisms (ORF>300 bp). The table does not include the 567,670-bp sequence of polydnavirus Cotesia congretata virus, because of its atypical low gene content (Espagne *et al.*, 2004). (1): URL: www.sanger.ac.uk/Projects/EhV/, (2): URL: pbi.bio.pitt.edu/. A: archea, B: proteobacteria;

| Species Name | Genome size (bp) | Domain | NCBI # |
|---|---|---|---|
| Canarypox virus | 359853 | Virus | NC_005309 |
| Coccolithovirus EhV-86 | 407,339 | Virus | 1 |
| Nanoarchaeum equitans | 490,885 | A | NC_005213 |
| Bacillus phage G | 497513 | Phage | 2 |
| Mycoplasma genitalium | 580,074 | B | NC_000908 |
| Buchnera aphidicola str. Bp | 615,980 | B | NC_004545 |
| Buchnera aphidicola str. Sg | 641,454 | B | NC_004061 |
| Buchnera aphidicola str. APS | 640,681 | B | NC_002528 |
| Wigglesworthia glossinidia | 697,724 | B | NC_004344 |
| Candidatus Blochmannia | 705,557 | B | NC_005061 |
| Ureaplasma parvum | 751,719 | B | NC_002162 |
| Mycoplasma mobile | 777,079 | B | NC_006908 |
| Mesoplasma florum | 793,224 | B | NC_006055 |
| Mycoplasma pneumoniae | 816,394 | B | NC_000912 |
| Onion yellows phytoplasma | 860,631 | B | NC_005303 |
| Mycoplasma hyopneumoniae | 892,758 | B | NC_006360 |
| Borrelia garinii | 904,246 | B | NC_006156 |
| Tropheryma whipplei | 927,303 | B | NC_004572 |
| Mycoplasma pulmonis | 963,879 | B | NC_002771 |
| Mycoplasma gallisepticum | 996,422 | B | NC_004829 |
| Chlamydia trachomatis | 1,042,519 | B | NC_000117 |
| Chlamydia muridarum | 1,072,950 | B | NC_002620 |
| Wolbachia endosymbiont strain TRS | 1,080,084 | B | NC_006833 |
| Rickettsia prowazekii | 111,1523 | B | NC_000963 |
| Rickettsia typhi | 1,111,496 | B | NC_006142 |
| Treponema pallidum | 1,138,011 | B | NC_000919 |
| Chlamydophila abortus | 1,144,377 | B | NC_004552 |
| Chlamydophila caviae | 1,173,390 | B | NC_003361 |
| Mimivirus | 1,181,404 | Virus | NC_006450 |

## Results and Discussion

**Giant viruses: a discontinuity in the distribution of virus genome sizes**

As of May 17, 2005, 447 double stranded DNA virus complete genome sequences were available at the National Center for Biotechnology Information. This data is unevenly distributed, with a few virus clades accounting for a large proportion of the known genomes such the *Caudovirales* (bacteriophage with tails, 168 sequences), and 4 major vertebrate-infecting virus families: 121 *Papillomaviridae*, 39 *Herpesviridae*, 28 *Adenoviridae*, and 21 *Poxviridae* (including 2 insect-infecting entomopox viruses). On the other hand, many families have less than a handful of representatives. Two of these under-represented families are associated to very large genomes, namely the *Nimaviridae* (Shrimp white spot syndrome virus, 305 kb), the *Phycodnaviridae* (*Paramecium bursaria* Chlorella virus 1, 331 kb, and *Ectocarpus siliculosus* virus, 336 kb). A list of the top-20 largest sequenced viral genomes is given in Table 2. An immediate conclusion can be drawn from this simple list: it is that large-sized genomes are not specifically associated to a given virus family, host type, vector, or ecological niche. A very diverse assortment of virus infecting bacteria, invertebrate, vertebrate, algae, or amoeba is found among the top-sized genomes.

This lack of correlation prompted us to use an objective data-mining technique to search for other putative structures in the viral genome size distribution. A traditional heuristic approach is to use a 2-dimensional data representation such as shown in Fig. 1 A. Here, the genome sizes are plotted (using a logarithmic scale) against their rank values in the distribution (the largest genome being ranked 1, the second largest being ranked 2, …, etc). It is expected that properties computed on a homogeneous category of "objects" (for instance viruses obeying similar evolutionary constraints) would be smoothly distributed in such a graph. Fig.1 A clearly shows this not to be the case. From rank 100 to rank 10, the logarithm of the genome sizes follow a slowly increasing quasi-linear distribution, the extrapolation of which would predict the largest viral genome (i.e. rank 1) to be about 260 kb in length. This part of the curve corresponds to viruses of various interspersed families (mostly bacteriophages, baculoviruses, herpesviruses, and poxviruses, see URL: www.giantvirus.org) the genome sizes (in a range of 110kb to 244 kb) of which appear to results from a variation of similar evolutionary constraints. This region of the distribution might be defined as the one encompassing the "regular" large ds-DNA viruses.

Despite the logarithmic size scale, the distribution of the top 9 ranking viruses is then separated from the previous ones by a clear gap also coinciding with a large change in slope. With the caveat expressed below, we propose that such an abrupt change in the distribution might define a new type of "giant viruses", corresponding to genome sizes of 280 kb and larger. However, even within this group of giant viruses, Mimivirus still appear to stands out.

**Table 2. Top-20 virus genome sizes.**

« Giant » viruses, corresponding to the discontinuity in the distribution of genome sizes (Fig. 1A) are in grey.

| Name | Genome Size (kb) | Family | NCBI Reference |
|---|---|---|---|
| Mimivirus | 1,181.4 | Mimiviridae | NC_006450 |
| Bacillus phage G | 497.5 | Myoviridae | pbi.bio.pitt.edu/ |
| Coccolithovirus EhV-86 | 407.3 | Phycodnaviridae | www.sanger.ac.uk/Projects/EhV/ |
| Canarypox virus | 359.9 | Chordopoxvirinae | NC_005309 |
| Ectocarpus siliculosus virus | 335.6 | Phycodnaviridae | NC_002687 |
| Paramecium bursaria Chlorella virus 1 | 330.7 | Phycodnaviridae | NC_000852 |
| Shrimp white spot syndrome virus | 305.1 | Nimaviridae | NC_003225 |
| Fowlpox virus | 288.5 | Chordopoxvirus | NC_002188 |
| Pseudomonas phage phiKZ | 280.3 | Myoviridae | NC_004629 |
| Bacteriophage KVP40 | 244.8 | Myoviridae | NC_005083 |
| Pongine herpesvirus 4 | 241.1 | Herpesviridae | NC_003521 |
| Melanoplus sanguinipes entomopoxvirus | 236.1 | Entomopoxvirus | NC_001993 |
| Human herpesvirus 5 strain Merlin | 235.6 | Herpesviridae | NC_006273 |
| Bacteriophage Aeh1 | 233.2 | Myoviridae | NC_005260 |
| Amsacta moorei entomopoxvirus | 232.4 | Entomopoxvirus | NC_002520 |
| Human herpesvirus 5 strain AD169 | 230.3 | Herpesviridae | NC_001347 |
| Murid herpesvirus 1 | 230.3 | Herpesviridae | NC_004065 |
| Murid herpesvirus 2 | 230.1 | Herpesviridae | NC_002512 |
| Heliothis zea virus 1 | 228.1 | Baculoviridae-like | NC_004156 |
| Cowpox virus | 224.5 | Chordopoxvirus | NC_003663 |

Figure 1. **Size *vs*. rank plot of the genome size distribution.** A) Top-largest viral genome. A logarithmic scale is used on the genome size axis. The arrow indicates a discontinuity and abrupt slope transition in the distribution, for genome sizes over 280 kb. B) Comparison with the genome size distribution of cellular prokaryotes (top curve). The slope transition corresponds to the parasitic/symbiotic bacteria with genome sizes below 1.2 Mb.

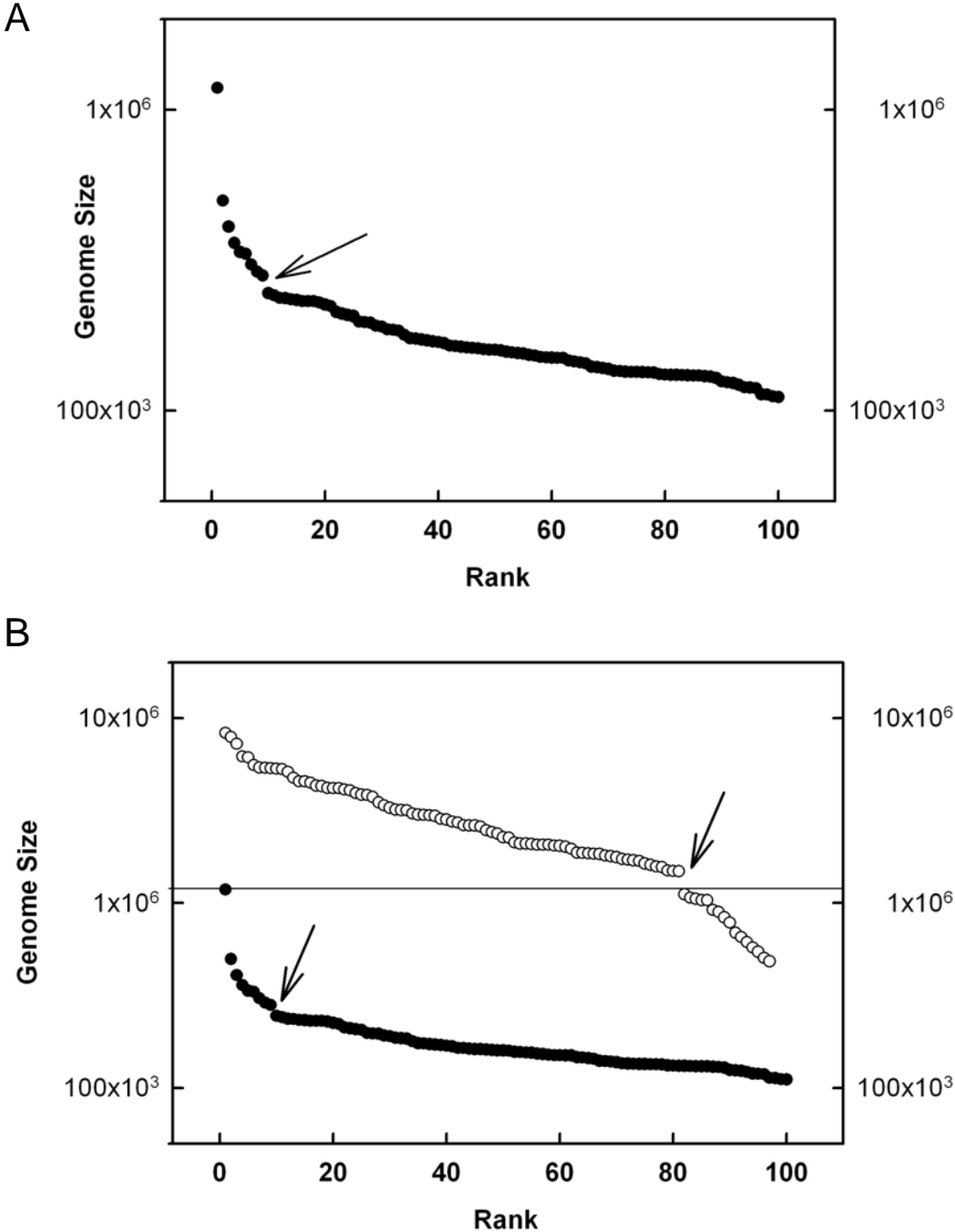

**Giant viruses encompass different families**

Interestingly, the select club of the giant viruses is no less diverse than the rest of the large ds-DNA viruses. Among its 9 members (see Table 2) one finds 2 poxviruses, 2 bacteriophages, 3 phycodnaviruses, the sole known Nimaviridae, and Mimivirus. This already suggests that, as for regular large ds-DNA viruses, genomic gigantism is not restricted to a specific host, a phylogenetic clade, or environmental niche. Various hypotheses can be proposed that might account for the observed discontinuity in the genome size distribution. One may propose, for instance, that en block genome duplication might have occurred independently in the various families resulting in sporadic super-sized members. Followed by the rapid functional diversification of the newly created paralogues, such an event might become evolutionarily advantageous, for instance by extending the host range (Mesyanzhinov et al., 2002). Despite a subsequent downsizing by gene loss, the genome size distribution might still reflect such discontinuous events. The discontinuity in the virus genome size distribution might also result from a trivial experimental bias. The change in slope might be interpreted as an indication that giant viruses (i.e. over 300 kb in genome size), although common in nature, might have a lesser probability to be isolated. At this point, one is struck by a coincidence: the genome size around which the discontinuity occurs corresponds to typical particle size of 200 nm in diameter. Sterilizing (i.e. bacteria removing) filters with 200nm to 300nm- diameter pore sizes might typically retain (or damage) viruses in this size range, eventually hindering their isolation or their serendipitous discovery. Such filters, invariably used in environmental sampling protocols designed to *a priori* separate cellular organisms from viruses, will also result in giant viruses material (e.g. DNA) to be misinterpreted as originating from cellular organisms (see the Metagenomics section). In conclusion, if the broken shape of the genome size distribution is due to this experimental bias, giant viruses might be much more frequent than presently thought. Many more giant viruses (including phages) might be still hiding themselves in the category of so-called "uncultivable" environmental bacteria.

**How big can a giant virus be?**

If Mimivirus and the other giant viruses are not as exceptional as suggested by the distribution in Fig 1. A, how big can a virus be? Are there natural limits to the particle and genome sizes of a virus, and what are they? In the case of Mimivirus, we already estimated that the central core region corresponded to a volume large enough to pack its DNA at a core concentration of about 450 mg/ml, a usual level encountered by other viruses (Raoult et al., 2004). Independent evaluations have proposed lower values and suggested that Mimivirus DNA density is in the lower range when compared to bacteriophages (Abrescia et al., 2004). Hence there is no evidence that Mimivirus particle size corresponds to a limit imposed by evolutionary or biophysical constraints. A 3-D reconstruction of the 190 nm-diameter PBCV-1 particle, the largest to date, fixed the precise number of major capsid protein molecules VP54 to 5040 (Nandhagopal et al., 2002). The twice as large (400 nm) Mimivirus particle should then be made of approximately 20,000 molecules of its major capsid protein. The Mimivirus particle is thus the largest (non-living?) nano-object capable of self-assembly. However, there are no obvious biophysical rules that would preclude even larger particles to exist.

In contrast, the typical linear dimension R of cellular organisms (e.g. bacteria) has to remain within a range dictated by the existence of a internal metabolic activity. This activity, roughly proportional to the cell volume (and thus to $R^3$) must be sustained by a flux of nutrient and energy dissipation through the surface of the cell envelope the area of which is proportional to $R^2$. This can be described by the qualitative equation: $\alpha.R^3 \approx \beta.R^2$, and thus forces R within the possible (µm) range of β/α (flux/metabolic activity).

In absence of metabolism, viral particles have no such constraints, and their volume are simply required to grow in direct proportion of the DNA to be packed in. The linear dimension (i.e. diameter) then only need to increase very slowly as the third root of the genome size. Thus, if Mimivirus can fit 1.2 Mb in a 0.4 µm-diameter particle, 6 Mb will fit in a 0.68 µm-diameter particle, and 10 Mb viral genome will only require a 0.8 µm-diameter particle.

Of course, a virus must remain small (< 1/30 of its host diameter for Mimivirus) compared to the dimension of the host it infects, and its genome size must remain in proportion to its host genome size (so that the host biosynthetic DNA machinery and nucleotide pool can suit the need of its replication). In that respect, bacteriophage G (genome size ~650kb, and 200 nm in diameter) (Serwer and Hayes, 2001; http://pbi.bio.pitt.edu/) infecting bacillus (with genome size of about 5 Mb, 2µm in size) may represent an extreme case. Accordingly, large amoebal protists could easily accommodate µm-sized, 10Mb-genome viruses, given their own enormous genome size (hundreds of Gbp) and cell dimension (150-4000 µm). Thus, we are forced to admit that the data at our disposal - possibly plagued by an experimental bias- do not suggest any clear limits on the genome size and complexity of the giant viruses that remain to be discovered.

**A comparison with the genome size distribution of prokaryotes**
Our assumption that the abrupt variation observed in the distribution of virus genome might have a biological significance can be somewhat validated by the results of the same analysis performed on cellular prokaryotes (Fig. 1 B). With the exception of a minor slope change for the 3 first ranking bacteria, the distribution exhibits a linear portion approximately covering the 1.5 Mb to 6 Mb genome size range, then a break clearly occurs for the lowest 16 ranking ones, pointing out bacteria exhibiting genomes smaller than extrapolated from their rank in the distribution. The main linear range encompasses all the regular "free living bacteria", the top 3 ranking genomes corresponding to the environmental "monsters", *Streptomyces avermitili*, *Streptomyces coelicolor*, *and Bradyrhizobium japonicum*, all with genome sizes larger than 9 Mb. More interestingly, the major distribution discontinuity nicely coincide with the genomes of parasitic/intracellular bacteria (Rickettsia, Buchnera, Mycoplasma, …, etc.), with genome smaller than 1.2 Mb. All these bacteria have undergone a similar genome reduction process while adapting to their host, a process made visible by our representation. Although we have no rationale explanation to offer, we noticed the striking coincidence of the discontinuity with the size of Mimivirus genome.

In conclusion, this application of the log(size) *vs*. rank plot to the cellular prokaryote genomes indicates: i) that the discontinuity exhibited in Fig.1A is not generated by the mathematical representation, ii) that such a discontinuity seems to coincide with the boundary between organisms submitted to broadly different evolutionary constraints (or isolation protocols).

**Mimicking a bacteria as a positively selected trait for amoebal virus?**
Mimivirus name originates from its initial misidentification as a Gram-positive bacteria, hence as a "microbe mimicking" virus. Two key factors were responsible for this mistake: a particle size allowing the virus to be easily visible with a light microscope, and its mild Gram-coloration (La Scola et al., 2003). These two properties might actually be central to the virus physiology. Electron microscopy study of amoeba, *Acanthamoeba polyphaga*, being infected by Mimivirus, strongly suggests that the virus is initially taken up via the feeding phagocytosis pathway normally used by the amoeba to feed on bacteria. It is known that the initial step of phagocytosis is more efficiently triggered by particle sizes in the µm range.

Using latex beads Korn and Weisman (1967) have shown that a transition toward a less active phagocytic behavior occurs for particles less than 0.6 µm in diameter. Interestingly, this is close to Mimivirus particle size. The formation of endocytic vesicles is also activated by the presence of a polysaccharide envelope on the engulfed particle, as typical for bacteria. This is mediated through the binding of amoebal lectin-type receptors. Electron microscopy and antibody binding patterns (La Scola, et al., 2005) strongly suggest that Mimivirus particles are actually encased in a 140-nm thick polysaccharidic layer, making it even more palatable for its amoebal host. This is consistent with the presence of many key sugar-manipulating enzymes encoded in the virus genome, some of them being quite specific of the biosynthesis of cell-surface lipopolysaccharide material such as perosamine (Raoult et al., 2004). The Gram staining of the virus is likely due to its LPS-like layer. This is also consistent with the extreme sturdiness of the particle as observed in our preliminary proteomics studies (Raoult et al., 2004). It is likely that Mimivirus is probably locked in this spore-like structure, and that the digestion of this LPS-like envelope by the amoeba endocytic vacuole is a prerequisite to the *bona fide* viral infection, that occurs through the vacuole membrane. The virus host specificity might be simply dictated by the presence or absence of the required enzymes in the phagosomes of various amoebal species (Weekers et al., 1995). In conclusion, the properties that led to the initial misinterpretation of Mimivirus as an amoeba-infecting bacteria, might actually be central to the life-style of many other giant viruses infecting their cellular host via the phagocytic route.

**Reassessing viral metagenomics: what is a virus sequence?**
As a consequence of the decreasing cost of sequencing DNA, the study of microbial biodiversity has now entered the genomic era, with the introduction of "metagenomics", defined as the culture-independent genomic analysis of an assemblage of microorganisms. Initial environmental sequencing projects targeted at 16S ribosomal RNA (rRNA) offered a glimpse into the phylogenetic diversity of uncultured organisms (reviewed in Riesenfeld et al., 2004; Delong 2005). The high-throughput shotgun sequencing of environmental DNA samples was then introduced to provide a more global views of those communities (Venter et al., 2004; Tringe et al., 2005). The metagenomic approach is now being specifically applied to the study of viral communities, using a gene-centric approach (Culley et al., 2003; Short and Suttle, 2002) or shotgun sequencing (Breitbart, 2002, 2003, 2004).

The existence of giant viruses comparable to small bacteria in terms of particle size and genome complexity makes the interpretation of metagenomic shotgun sequences much less straightforward than previously thought, on two counts. First, filtering steps are invariably used to separate the "bacterial" fraction (using a 0.1 to 0.3 µm pore size range) from the "viral" fraction. In consequences, non-filtering giant viruses will contribute sequences misinterpreted as part of the bacterial pool, while they will be missing from the survey of viral communities. Given the tendency of algal viruses to be large (e.g. phycodnaviruses, Van Etten et al., 1991; Van Etten, 2003; Claverie, 2005) and the fact that viruses might outnumber bacteria by an order of magnitude in some aquatic environments (Wommack et al., 1992; Wommack and Colwell, 2000), the results from these ecological surveys should be interpreted with caution. For instance, an exhaustive similarity search (Ghedin and Claverie, 2005) of all Mimivirus predicted proteins against all publicly available sequences identified many of their closest homologues among the "bacterial" pool of the Sargasso Sea environmental sequences (Venter et al., 2004). More detailed phylogenetic analyses strongly suggested that these environmental sequences do belong to unknown large viruses evolutionarily closer to Mimivirus than to any presently characterized viral species (Ghedin and Claverie, 2005).

A second -more fundamental- problem is that, in absence of the simple "filtering criteria", distinguishing giant virus genes from those found in cellular organisms becomes very tricky, when solely based on sequence similarity. A funny example is given by the presence of an acetylcholinesterase-like gene in *Acanthamoeba polyphaga* Mimivirus. ORF L906 is 1,737 bp-long encoding a 579-residue putative protein. The best BLASTP (Altschul et al., 1997) matching homologue of Mimivirus L906 in the nr database is *Torpedo californica* acetylcholinesterase (GenBank accession number CAA27169) exhibiting 30.1 % identity over 508 residues. No viral sequence exhibits any significant similarity with Mimivirus L906. Captured in a metagenomic sampling, such a sequence will no doubt be classified as originating from a cellular organism, if not from a contamination from fish DNA! Yet this sequence is part of Mimivirus genome. Incidentally, many L906 homologues are found in the Sargasso Sea environmental data set, most of them from bacterial origin (probably).

The multi-alignment of MIMI_L906 (using T-Coffee, Poirot et al., 2004) with a set of carboxylesterases 3-D structures and typical acetylcholinesterase sequences (Fig. 2) shows the presence of the expected catalytic triad involving a serine, a glutamate (or aspartate) and a histidine, as well as other motifs (Krejci et al., 1991). A phylogenetic analysis using MEGA3 (Kumar et al., 2001) clustered Mimivirus L906 with prokaryotic paranitrobenzyl carboxylesterases (Fig. 3) that are known to catalyze the hydrolysis of the para-nitrobenzyl esters of various β-lactam antibiotics. Although the exact role of the enzyme in Mimivirus is not known, it might play a role in the disruption of the amoeba phagocytic vacuole membrane, helping the virus to gain access to the host cytoplasm.

**Translation apparatus genes: the final frontier between cells and viruses?**

With more bacterial genomes being sequenced, the number of core genes strictly shared by all bacterial or archebacterial species, including the smallest parasitic ones, has been steadily decreasing. It is now down to 60 genes, including ribosomal proteins, aminoacyl-tRNA synthetases, and the core components of the transcription and DNA replication apparatus (Crapoulet et al., 2005). This can be summarized by saying that parasitic bacteria have found evolutionary solutions to eventually dispense with most functions, but those encoded by these 60 "core" genes. Of course, some of these cellular genes reputed irreplaceable today, might be found absent from a microbial genome sequenced in the future. Already, the nucleomorph (i.e. the enslaved algal nucleus) of the cryptomonad alga *Guillardia theta* does not appear to encode a complete set of ribosomal proteins, and might thus have developed an import mechanism for them (Douglas et al., 2001).

On the other hand, giant virus genomes exhibit an increasingly large assortment of biosynthetic pathways and regulatory components, and most of them exhibit their own DNA replication and transcription apparatus. The variability of their gene contents is so large that any type of gene might eventually turn out in the next giant virus genome that will be sequenced.

Figure 2. **Comparison of Mimivirus L906 putative protein sequence with typical carboxylesterase amino acid sequences, including proteins of known 3-dimentional structures.** The multiple alignment was generated with 3D-coffee (Poirot et al., 2004). The highly conserved catalytic triad is composed of a serine at position 200, a glutamate at position 331, and a histidine at position 444. Critical/conserved residues are boxed. Approximately 40 C-terminal residues exhibiting little conservation across the different sequences are omitted.

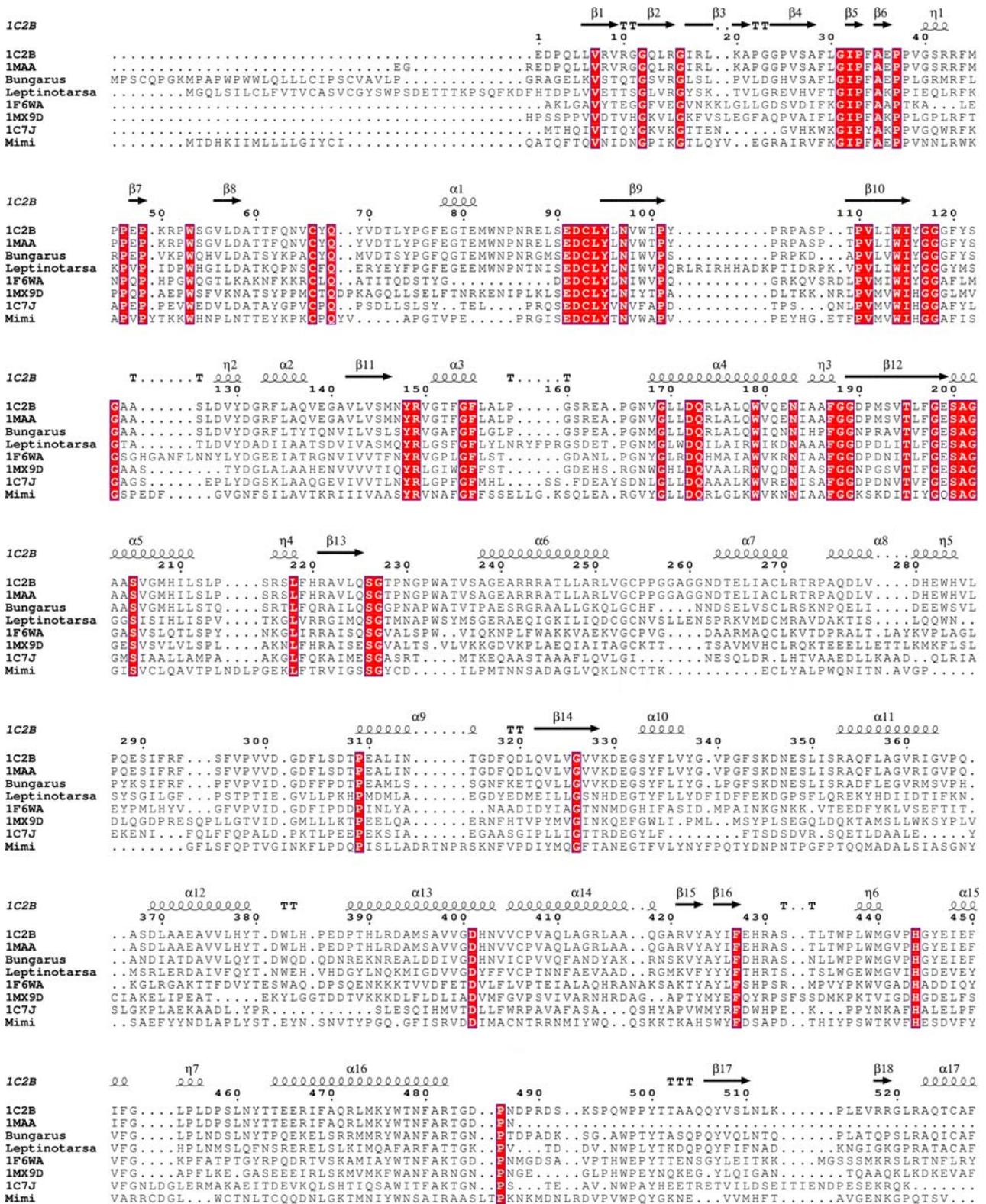

**Figure 3. Phylogenetic classification of the Mimivirus L906 encoded putative carboxylesterase among type B carboxylesterases and other esterases/lipases.** The Neighbor-joining method with poisson correction as provided by the MEGA3 software (Kumar et al., 2001) was used. Bootstrap values above 50% are indicated.

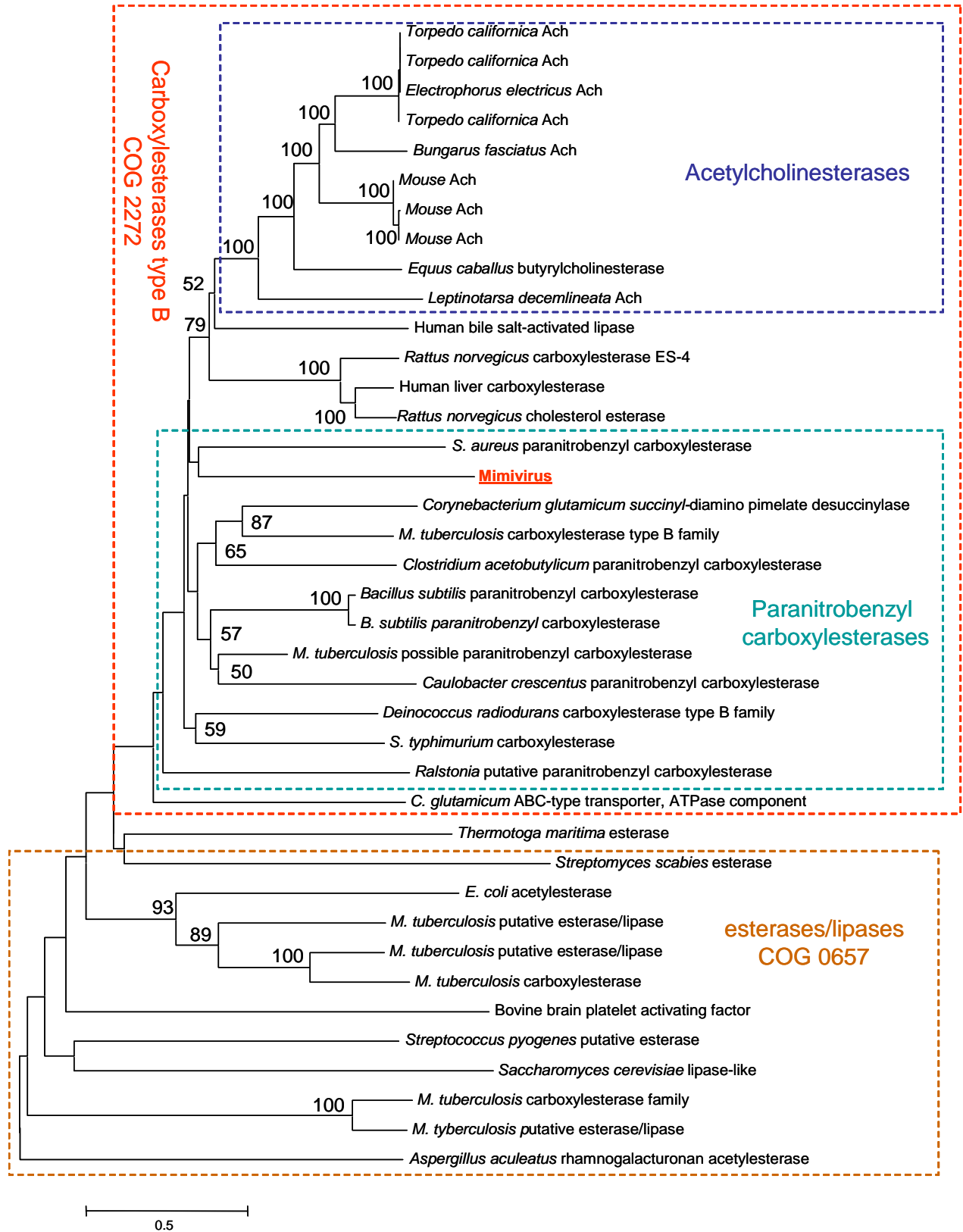

The difference that was thought to unequivocally separate the cellular world from the one of viruses was the presence of a working translation apparatus. However, following the presence of tRNAs in the genome of many giant viruses, the discovery of many translation factors and several aminoacyl-tRNA synthetases encoded in Mimivirus genome was a serious blow to this last resort criteria. By now, finding several ribosomal protein genes - or even a ribosomal RNA-like sequence -, in a future giant virus genome becomes much less unexpected.

With its size and genome complexity for the first time overlapping with those of cellular organisms, Mimivirus shattered at once a century old portrait of what a *bona fide* virus should look like. This, however, is not a simple epistemological readjustment of no practical consequences. Mimivirus is probably the first representative of a long list of many more giant viruses, the proper classification of which will pose a serious and durable challenge to our definition of life forms. As an immediate concern, the interpretation of metagenomic data must now been reappraised. If there is no single gene that is common to all viral genomes (Edwards and Rohwer, 2005), any gene might also eventually turn out in a giant virus. In consequence, the attribution of remotely similar environmental sequences to a giant virus or a cellular organisms can not simply rely on the identity of its closest homologue, but must include a complete taxonomic assessment (e.g. using the best reciprocal match criteria for orthology) followed by a detailed phylogenetic analysis (Ghedin and Claverie, 2005). Even though, a doubt might still subsist about the origin of these sequences, in the case of low sequence similarity and/or insufficient coverage of certain clades.

**The mysterious origin of giant viruses**
The evolutionary forces at the origin of giant viruses (loosely defined as those packing more than 300 genes in 200-300 nm-diameter particles) and the rationale behind their genome complexity are not understood. Various hypotheses can be proposed, from traditional to the most revolutionary. Prior to expose some of them, it is worth to notice that these viruses are truly more complex than their leaner counterparts (e.g. the typical 50kb genome adenoviruses or phages): the increase in genome size is not due, for instance, to the accumulation of non-coding repeats, junk DNA, or the huge expansion of a few gene families. Their ultrastructure also appears more complex than their smaller relatives, as confirmed by direct proteomic analyses (Raoult et al., 2004). Nothing, in the recognizable gene content of giant viruses, appears to predispose them to the capture and accumulation of random DNA segments: in contrast with promiscuous bacteria, their genome is not particularly enriched with mobile elements, palindromic structures, or genes encoding the necessary enzymatic equipment (such as transposases, integrases, …, etc.).

Yet, viruses are traditionally seen as being prone to frequent lateral gene transfer from their host, and their genomes are considered like bags of genes more or less randomly accumulated foreign genes, around the limited set of conserved core genes (Iyer et al., 2001) pertaining to essential functions.

We have shown elsewhere (Ogata et al., 2005b) that this picture, inherited from the world of retroviruses and transducing phages, does not agree with our analysis of Mimivirus genome: a small proportion (40%) of its predicted proteins does exhibit a significant similarity within the sequence databases, and even less (1%) exhibit their best matches against recently determined *Entamoeba histolytica* genome sequence (Ogata et al., 2005b) or against *Acanthamoeba castellani* (a close relative of Mimivirus host amoeba) sequence data (Fig. 4).

Figure 4. **Distribution of BLAST scores of Mimivirus ORFs against *Acanthamoeba castellani* sequences** (vertical axis; TBLASTN) and non-protozoa sequences (horizontal axis; BLASTP). The *A. castellani* sequence set is composed of sequences downloaded through the NCBI Entrez system (18,433 sequences; 19.3 Mbp in total) and those obtained through the Protist EST Program (http://megasun.bch.umontreal.ca/pepdb/pep_main.html; 5,243 sequences; 2.5 Mbp in total). BLASTP scores against non-protozoa sequences were obtained using partial Mimivirus sequences (HSPs matching to *A. castellani* data). Abbreviations are as follows: dTDP4DR (dTDP-4-dehydrorhamnose reductase); HSP70 (70-kDa heat-shock protein); RNRL (ribonucleotide reductase large subunit); RpbL (RNA polymerase II largest subunit).

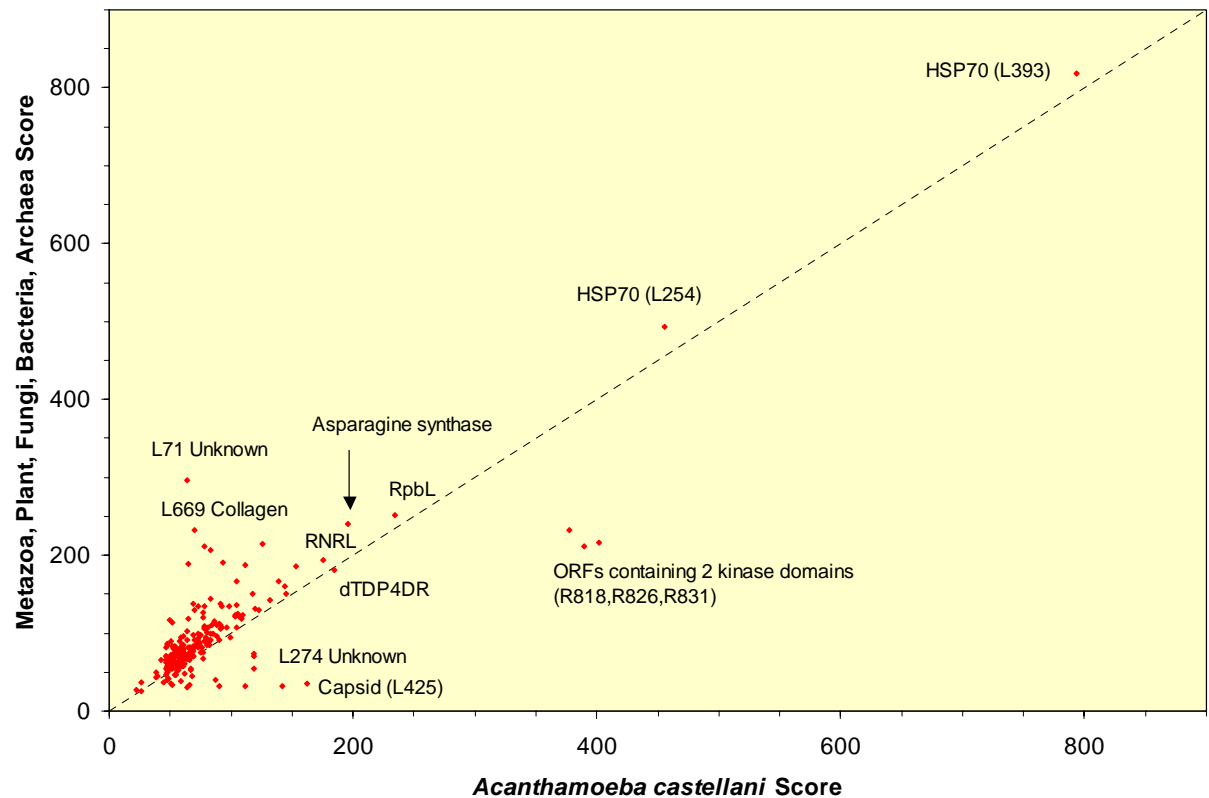

Using a Bayesian method based on nucleotide word frequencies (Nakamura et al., 2004), we estimated that less than 9% of Mimivirus gene might have been recently acquired by lateral gene transfer. Our studies based on phylogenetic tree inference suggest several putative horizontal gene transfers (http://www.giantvirus.org/mimitrees/). For instance, the Mimivirus mismatch repair ATPase (MutS) is most closely related to a homologue encoded in the mitochondrial genome of *Leptogorgia chilensis*. However, such cases apparently involving horizontal gene transfer were limited. Other arguments, such as the highly conserved structure of the promoter regions of Mimivirus genes (49% of them exhibiting a unique and strictly conserved motif) (Suhre et al., submitted paper available on arXiv: http://arxiv.org/abs/q-bio.GN/0504012) militate against a mosaic structure of Mimivirus genome and a high prevalence of horizontal gene transfer. Thus, this mechanism is not responsible for the huge increase of Mimivirus gene content compared to regular DNA viruses.

Another striking feature of Mimivirus genome is the absence of the sign of reductive genome evolution or pseudogenes. The genome is packed with genes (with an average intergenic region size of 150 nt), none of them showing any indication of degradation. Figure 5 shows that Mimivirus genome exhibit a higher ORF density for any size ranges than the genomes of *Rickettsia prowazekii* or *Mycobacterium leprae* known to contain a large proportion of "junk" DNA and pseudo-genes due to ongoing reductive evolution (Fig. 5). The Mimivirus ORF density is close to that of *Escherichia coli*, with a higher proportion of ORF larger than 400 residues. The absence of decaying genes argues against Mimivirus large genome being the result of frequent acquisitions of more or less random host genes. It is also in contrast with the clear tendency toward genome reduction/gene degradation observed for most intracellular parasitic bacteria, with similar or smaller genome sizes (such as *Rickettsia*, *Buchnera*, or *Mycoplasma*, Fig. 1 B) (Ogata et al., 2001; Andersson and Andersson, 2001; Moran, 2002).

In contrast to these cellular organisms appearing to be irreversibly evolving toward an increasingly host-dependent life-style, Mimivirus and other giant viruses appear to be in an evolutionary steady state, showing no tendency toward reducing their size. On the contrary, Mimivirus exhibits some large families of paralogues originating from relatively recent multiplication/duplication events (Suhre et al., submitted paper available at arXiv: http://arxiv.org/abs/q-bio.GN/0505049). A residual signal from a more ancient segmental genome duplication event of about 200,000 bp can also be detected. Some of the larger families of recently duplicated genes correspond to tandem duplications of up to 11 copies in a row (genes L175 to L185). A phylogenetic analysis of these genes indicates that they are not the result of a burst multiplication of the same ancestral gene, but that they were derived from distinct duplication events, and evolved independently following their creation. These events may thus be regarded as neo- or sub-functionalization events. Using remote homology detection methods (Soding, 2004), a number of these gene families can be linked to functions such as transcription control, cell signaling and protein ubiquitination. We therefore speculate that these genes may play a role in recently acquired and/or diversified host adaptation functions. If any, Mimivirus genome shows more signs toward expansion than to reduction!

Figure 5. **Comparison of ORF size cumulative distributions.** The graph shows the numbers of ORFs (y-axis) (per 1 Mb), that are longer than or equal to X-codons (x-axis) annotated in the genomes of Mimivirus, *Escherichia coli*, *Rickettsia prowazekii* and *Mycobacterium leprae*.

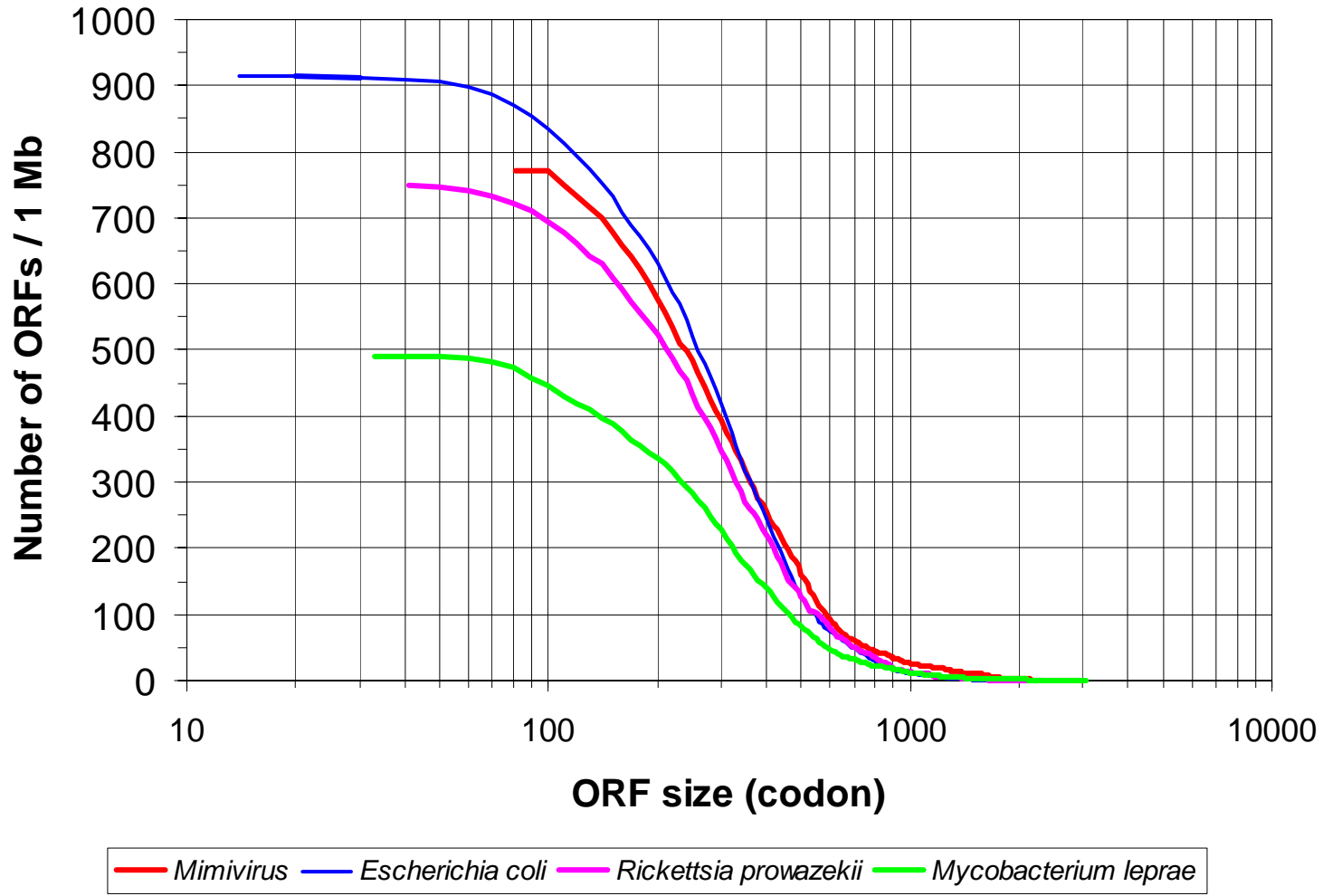

At variance with the classical "incremental bag of genes" view of DNA virus, we have proposed (Raoult et al. 2004; Ogata et al., 2005b) that the Mimivirus lineage might have emerged before the individualization of cellular organisms from the three domains of life prior to the Darwinian threshold (Woese, 2002), eventually participating to the mixing of bacterial and archeal genes that led to the emergence of the eukaryotic cell (reviewed in Pennisi, 2004; Ogata et al., 2005b). In that context, different viral genes are not expected to exhibit entirely consistent phylogenies, and the similarity between extant orthologues in cellular or other viral species is expected to be very low. Clearly, more genome sequences of giant viruses, functional genomics studies, and the determination of the 3-D structures of many of the viral gene products bearing no sequence homology to any other proteins, are needed to better understand the origin of these giant viruses, or "girus".